\documentclass[
aps,epsf,
superscriptaddress,
twocolumn,
amsmath,amssymbol,
]
{revtex4-1}

\usepackage[dvipdfmx]{graphicx}
\usepackage[dvipdfmx]{color}

\begin{document}


\title{Muon spin relaxation study of the spin correlation in the overdoped regime of electron-doped high-$T_{\rm c}$ cuprate superconductors}

\author{Malik A. Baqiya}
\affiliation{Department of Physics, Faculty of Mathematics and Natural Sciences, Institut Teknologi Sepuluh Nopember (ITS), ITS Campus, Sukolilo, Surabaya 60111, Indonesia}
\author{Tadashi Adachi}
\email[Corresponding author: ]{t-adachi@sophia.ac.jp}
\affiliation{Department of Engineering and Applied Sciences, Sophia University, 7-1 Kioi-cho, Chiyoda-ku, Tokyo 102-8554, Japan.}
\author{Akira Takahashi}
\affiliation{Department of Applied Physics, Tohoku University, 6-6-05 Aoba, Aramaki, Sendai 980-8579, Japan}
\author{Takuya Konno}
\affiliation{Department of Applied Physics, Tohoku University, 6-6-05 Aoba, Aramaki, Sendai 980-8579, Japan}
\author{Taro Ohgi}
\affiliation{Department of Applied Physics, Tohoku University, 6-6-05 Aoba, Aramaki, Sendai 980-8579, Japan}
\author{Isao Watanabe}
\affiliation{Meson Science Laboratory, Nishina Center for Accelerator-Based Science, The Institute of Physical and Chemical Research (RIKEN), 2-1 Hirosawa, Wako 351-0198, Japan}
\author{Yoji Koike}
\affiliation{Department of Applied Physics, Tohoku University, 6-6-05 Aoba, Aramaki, Sendai 980-8579, Japan}

\date{\today}

\begin{abstract}
In order to investigate the low-energy antiferromagnetic Cu-spin correlation and its relation to the superconductivity, we have performed muon spin relaxation ($\mu$SR) measurements using single crystals of the electron-doped high-$T_{\rm c}$ cuprate Pr$_{1-x}$LaCe$_x$CuO$_4$ in the overdoped regime. 
The $\mu$SR spectra have revealed that the Cu-spin correlation is developed in the overdoped samples where the superconductivity appears. 
The development of the Cu-spin correlation weakens with increasing $x$ and is negligibly small in the heavily overdoped sample where the superconductivity almost disappears. 
Considering that the Cu-spin correlation also exist in the superconducting electron-doped cuprates in the undoped and underdoped regimes [T. Adachi {\it et al.}, J. Phys. Soc. Jpn. {\bf 85}, 114716 (2016)], our findings suggest that the mechanism of the superconductivity is related to the low-energy Cu-spin correlation in the entire doping regime of the electron-doped cuprates.
\end{abstract}

\maketitle

\section{Introduction}\label{sec:introduction}
In the research of high-$T_{\rm c}$ cuprate superconductivity, the relationship between the Cu-spin correlation and superconductivity has been the central issue in both hole-doped and electron-doped cuprates. 
For the hole-doped cuprate of La$_{2-x}$Sr$_x$CuO$_4$ (LSCO), neutron-scattering experiments have revealed that the commensurate Cu-spin correlation in the antiferromagnetic (AF) state of the parent compound changes to the incommensurate one with hole doping in the superconducting (SC) state, \cite{yamada-prb} followed by the disappearance of both incommensurate Cu-spin correlation and superconductivity in the heavily overdoped regime. \cite{wakimoto} 
Muon-spin-relaxation ($\mu$SR) measurements in Zn-impurity-substituted La$_{2-x}$Sr$_x$Cu$_{1-y}$Zn$_y$O$_4$ have revealed that the development of the Cu-spin correlation vanishes at the end point of the SC region in the heavily overdoped regime of the phase diagram. \cite{risdi-lsco} 
Therefore, the incommensurate Cu-spin correlation appears to be intimately related to the superconductivity. 
For the electron-doped cuprates, on the other hand, the commensurate Cu-spin correlation has been observed in the optimally doped regime of Nd$_{2-x}$Ce$_x$CuO$_4$ \cite{yamada-prl} and Pr$_{1-x}$LaCe$_x$CuO$_4$ (PLCCO). \cite{kang,wilson} 
The relationship between the commensurate Cu-spin correlation and superconductivity has been unclear in the electron-doped cuprates.

Recently, the so-called undoped (Ce-free) superconductivity in the electron-doped cuprates has attracted considerable research attention. 
It has been reported that the superconductivity appears even in the parent compound of $x=0$ and in a wide range of $x$ in Nd$_{2-x}$Ce$_x$CuO$_4$ thin films through the appropriate reduction annealing to remove excess oxygen from the as-grown thin films. \cite{tsukada,matsumoto} 
The superconductivity in the parent compound has also been confirmed in the polycrystalline samples. \cite{asai,takamatsu} 
Two possible mechanisms of the undoped superconductivity have been proposed; the electron doping by the oxygen deficiency (oxygen non-stoichiometry) \cite{horio-nco} and the collapse of the charge-transfer gap due to square-planer coordination of oxygen in the CuO$_2$ plane. \cite{adachi-jpsj} 
If the latter is the case, the undoped superconductivity indicates that the phase diagram is completely different from the former one, that is, the superconductivity in the electron-doped cuprates cannot be understood in terms of carrier doping into the parent Mott insulators as in the case of the hole-doped cuprates. 
An important issue is whether or not the Cu-spin correlation is related to the superconductivity in the electron-doped cuprates. 

Through the improved reduction annealing, high-quality SC single crystals have been obtained in underdoped Pr$_{2-x}$Ce$_x$CuO$_4$ with $x \ge 0.04$ \cite{brinkmann} and Pr$_{1.3-x}$La$_{0.7}$Ce$_x$CuO$_{4+\delta}$ with $x \ge 0.05$. \cite{adachi-jpsj,horio,adachi-review} 
Formerly, we have performed $\mu$SR measurements of the SC parent polycrystal of La$_{1.8}$Eu$_{0.2}$CuO$_4$ and the SC underdoped single crystal of Pr$_{1.3-x}$La$_{0.7}$Ce$_x$CuO$_4$ with $x=0.10$. \cite{adachi-review,adachi-jpsj2} 
It has been found that a short-range magnetic order is formed at low temperatures in both samples, suggesting a coexisting state of superconductivity with the short-range magnetic order. 
The development of the Cu-spin correlation has also been confirmed in $\mu$SR measurements of the SC parent thin film of La$_{1.9}$Y$_{0.1}$CuO$_4$. \cite{kojima} 
These results suggest that a small amount of residual excess oxygen in a sample causes the development of the Cu-spin correlation and/or the formation of the short-range magnetic order, indicating a strongly correlated electron system of the undoped and electron-underdoped cuprates.

The next issue is how the Cu-spin correlation changes with electron doping concomitant with the weakening of the superconductivity in the overdoped regime. 
Inelastic neutron-scattering experiments in the overdoped PLCCO with $x \le 0.18$ have revealed that the characteristic energy of the Cu-spin correlation decreases with increasing $x$ and seems to disappear with the superconductivity. \cite{fujita} 
This is different from the results of the hole-doped cuprates in which the characteristic energy of the Cu-spin correlation is unchanged but the spectral weight decreases with hole doping, \cite{wakimoto} suggesting the occurrence of a phase separation into SC and normal-state regions in a sample. \cite{tanabe} 
From the former $\mu$SR measurements in the SC polycrystal of PLCCO with $x=0.14$, slowing down of the Cu-spin fluctuations has been observed at low temperatures without any magnetic order. \cite{risdi-plcco} 
NMR experiments of the SC single crystal of Pr$_{1.3-x}$La$_{0.7}$Ce$_x$CuO$_4$ with $x=0.15$ have also indicated the presence of AF spin fluctuations. \cite{yamamoto} 
These suggest that, compared with the short-range magnetic order in the parent and underdoped samples, \cite{adachi-review,adachi-jpsj2} the development of the Cu-spin correlation weakens with increasing $x$ but is apparently observed in the slightly overdoped regime. 
In order to obtain detailed information on the low-energy Cu-spin correlation in the heavily overdoped regime and its relation to the superconductivity, we have carried out $\mu$SR measurements using PLCCO single crystals in the heavily overdoped regime of $x=0.17$ and $0.20$.

\section{Experimental}
Single crystals of PLCCO with $x=0.17$ and $0.20$ were prepared by the traveling solvent floating zone method. \cite{lambacher,malik} 
The quality of the grown crystals was checked by the x-ray back-Laue photography and powder x-ray diffraction to be good. 
The composition of the crystals was analyzed by the inductively-coupled-plasma spectrometry. 
For the reduction annealing in a vacuum condition of $2 \times 10^{-4}$ Pa, the two-step annealing was performed at 900$^{\rm o}$C for 12 h and 500$^{\rm o}$C for 12 h for $x=0.17$. 
For $x=0.20$, the improved one-step reduction annealing was carried out at 800$^{\rm o}$C for 24 h. \cite{adachi-jpsj} 
Magnetic-susceptibility measurements were performed using a SC quantum interference device (SQUID) magnetometer (Quantum Design, MPMS). 
Figure 1 shows the temperature dependence of the magnetic susceptibility of PLCCO with $x=0.17$ and $0.20$ together with $x=0.13$ and $0.15$. \cite{malik} 
The SC transition temperature $T_{\rm c}$ of $x=0.17$ is $\sim 5$ K and the Meissner diamagnetism at 2 K is much smaller than those of $x=0.13$ and $0.15$, indicating that the superconductivity is weak. 
For $x=0.20$, the Meissner diamagnetism is unobservable, indicating a non-SC state of this sample. 
As shown in the inset of Fig. 1, values of $T_{\rm c}$ are almost consistent with those in the former report. \cite{fujita} 
Zero-field (ZF) and longitudinal-field (LF) $\mu$SR measurements were performed at low temperatures down to 0.3 K at the RIKEN-RAL Muon Facility at the Rutherford-Appleton Laboratory in the United Kingdom using a pulsed positive surface muon beam. 
The data were analyzed using the WiMDA program. \cite{pratt}

\section{Results and Discussion}
\begin{figure}[tbp]
\begin{center}
\includegraphics[width=1.0\linewidth]{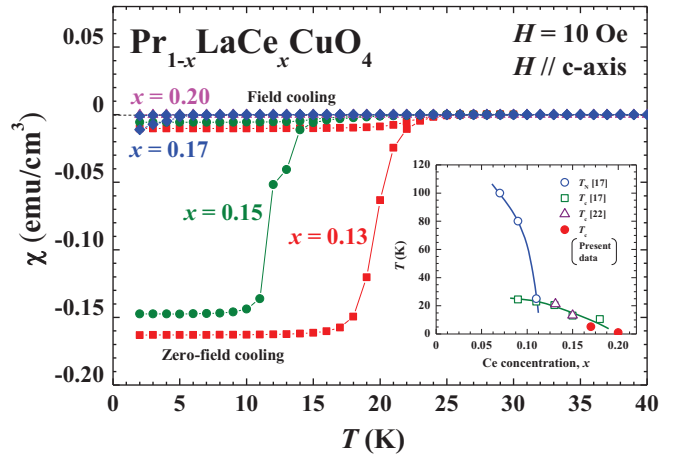}
\caption{(Color online) Temperature dependence of the magnetic susceptibility of reduced single crystals of PLCCO with $x=0.17$ and 0.20 together with $x=0.13$ and 0.15. \cite{malik}  The inset shows the Ce-concentration dependence of $T_{\rm c}$ in PLCCO, together with the former results of $T_{\rm c}$ and $T_{\rm N}$. \cite{fujita}  Solid lines are to guide the reader's eye}
\label{fig:Figure1}
\end{center}
\end{figure}

Figure 2 shows the ZF-$\mu$SR time spectra of the reduced crystals of PLCCO with $x=0.17$ and $0.20$ together with $x=0.14$. \cite{risdi-plcco}
In both $x=0.17$ and $0.20$, the depolarization of muon spins is slow at a high temperature of 200 K owing to the very small nuclear dipole field randomly oriented at the muon site, indicating an almost paramagnetic state of Cu spins. 
With decreasing temperature, the depolarization of muon spins becomes fast as seen in $x=0.14$ due to static random magnetism of small magnetic moments of Pr$^{3+}$ ions induced by the mixing of the excited state in the crystal electric field. \cite{risdi-plcco,kadono}

For $x=0.14$, the development of the Cu-spin correlation is characterized by the increase in the asymmetry in the long-time region above $\sim 4$ $\mu$sec with decreasing temperature shown in Fig. 2(a), \cite{risdi-plcco} which is due to the recovery of the asymmetry toward 1/3 in a magnetically ordered state. 
It is found that the recovery of the asymmetry in the long-time region is negligibly small at low temperatures down to 9 K for $x=0.17$ and down to 0.3 K for $x=0.20$. 
This indicates that the Cu-spin correlation is hardly developed in the heavily overdoped regime of PLCCO where the superconductivity almost disappears.
 
To see effects of the Cu-spin correlation in detail, the ZF-$\mu$SR time spectra were analyzed using the following two-component function, \cite{risdi-plcco}
\begin{equation}
A(t) = A_{\rm s} {\rm exp}[-(\lambda t)^\beta] + A_{\rm G} {\rm exp}[-\sigma^2 t^2] + A_{\rm base}.
\label{eq1} 
\end{equation}
The first term represents a stretched-exponential component in which effects of nuclear spins and Cu spins are dominant. 
The $A_{\rm s}$, $\lambda$, and $\beta$ are the initial asymmetry, depolarization rate of muon spins, and power of damping, respectively. 
The second term represents a static Gaussian component in which the effect of small Pr$^{3+}$ moments is dominant. 
The $A_{\rm G}$ and $\sigma$ are the initial asymmetry and depolarization rate of muon spins, respectively. 
The $A_{\rm base}$ is a time-independent background term.
The spectra are well fitted with Eq. (\ref{eq1}), as clearly shown in Figs. 2(b) and 2(c). 
It is noted that the use of the two terms with $A_{\rm s}$ and $A_{\rm G}$ suggests possible two muon stopping sites in PLCCO.
Although former reports have suggested one muon stopping site in the T'-cuprates based on the dipole-field calculation, \cite{luke,le} a recent first-principle calculation suggests two muon stopping sites; one is near the CuO$_2$ plane mainly sensing the dipole field of the Cu spins and the other is near the (Pr,La,Ce)-O layer mainly sensing that of Pr$^{3+}$ moments. \cite{tsutsumi,tsutsumi-calc}

\begin{figure}[tbp]
\begin{center}
\includegraphics[width=1.0\linewidth]{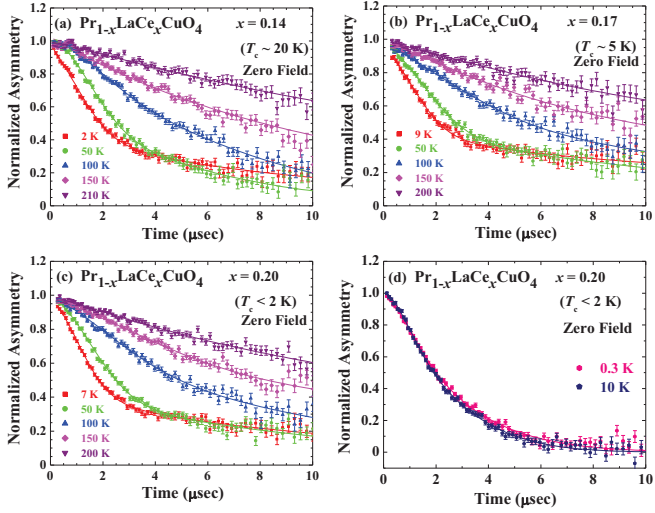}
\caption{(Color online) ZF-$\mu$SR time spectra of reduced PLCCO crystals with (a) $x=0.14$, \cite{risdi-plcco} (b) $x=0.17$, and (c) $x=0.20$ at various temperatures.  (d) ZF-$\mu$SR time spectra of the reduced PLCCO crystal with $x=0.20$ at 10 K and 0.3 K.  Solid lines are the best-fit results using the two-component function described in Eq. (1).}
\label{fig:Figure2}
\end{center}
\end{figure}

Figure 3 shows the temperature dependence of the fitting parameters $A_{\rm s}$, $\beta$, and $\sigma$, and $\lambda$ for both crystals of PLCCO with $x=0.17$ and $0.20$ together with $x=0.14$. \cite{risdi-plcco} 
At high temperatures above 100 K, all parameters seem to be almost independent of temperature and the normalized $A_{\rm s}$ is nearly one. 
It is noted that the change of the spectra above 100 K shown in Fig. 2 for all samples is predominantly due to the small change of $A_{\rm s}$.
The $\sigma$ ($A_{\rm s}$) increases (decreases) with decreasing temperature below $\sim 100$ K owing to the growing effect of Pr$^{3+}$ moments. \cite{risdi-plcco} 
For $x=0.14$, the development of the Cu-spin correlation is characterized by the steep increase in $\lambda$ and enhancement of $A_{\rm s}$ below $\sim 30$ K as shown in Figs. 3(d) and 3(a). \cite{risdi-plcco} 
For $x=0.17$ and $0.20$, on the contrary, neither steep increase in $\lambda$ nor apparent enhancement of $A_{\rm s}$ is observable at low temperatures, indicating that the development of the Cu-spin correlation is negligibly small in both samples.

\begin{figure}[tbp]
\begin{center}
\includegraphics[width=0.8\linewidth]{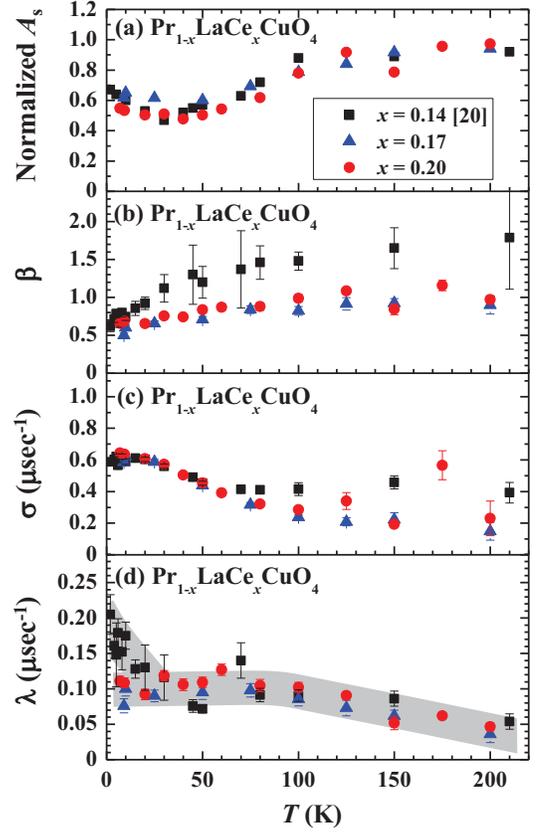}
\caption{(Color online) Temperature dependence of the fitting parameters in the two-component function described in Eq. (1) for reduced crystals of PLCCO with $x=0.17$ and 0.20 together with $x=0.14$ \cite{risdi-plcco} for comparison.}
\label{fig:Figure3}
\end{center}
\end{figure}

In order to further investigate the effects of Cu spins and Pr$^{3+}$ moments, LF-$\mu$SR measurements were performed under LF up to 1000 G at 0.3 K for PLCCO with $x=0.20$. 
As shown in Fig. 4, the tail of the spectrum is gradually quenched with increasing field up to 100 G. 
This suggests the existence of static magnetism due to Pr$^{3+}$ moments. 
In the long-time region, the slow depolarization is still observed up to 1000 G, indicating that there exist fluctuating internal fields at the muon site due to Cu spins. \cite{risdi-plcco} 
Therefore, the LF-$\mu$SR results suggest the coexistence of the static magnetism of Pr$^{3+}$ moments and fluctuating Cu spins in $x=0.20$ as well as $x=0.14$. \cite{risdi-plcco}

For both PLCCO crystals with $x=0.17$ and $0.20$, the $\mu$SR spectra show the existence of both the static magnetic field due to Pr$^{3+}$ moments and the slow depolarization related to Cu-spin fluctuations. 
The development of the Cu-spin correlation becomes weak with increasing $x$ and is negligibly small at $x=0.20$ where the superconductivity almost disappears. 
Combined with the results in the undoped and underdoped regimes, \cite{adachi-review,adachi-jpsj2} these results suggest an intimate relationship between the Cu-spin correlation and superconductivity in the entire doping regime of PLCCO. 

\begin{figure}[tbp]
\begin{center}
\includegraphics[width=1.0\linewidth]{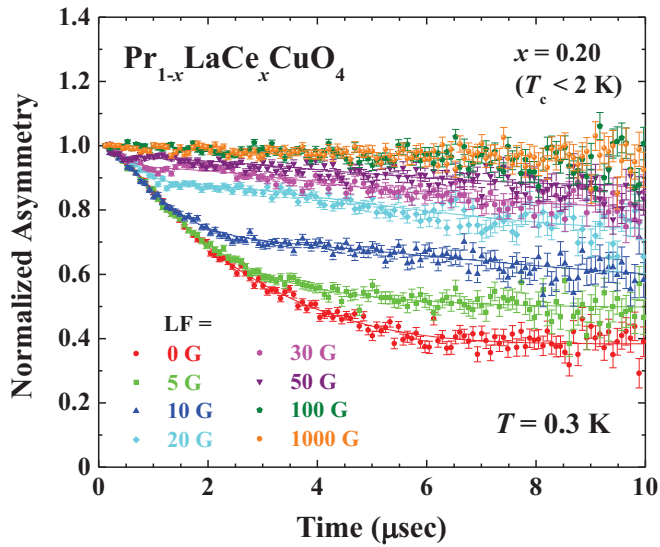}
\caption{(Color online) LF-$\mu$SR time spectra of the reduced crystal of PLCCO with $x=0.20$ at 0.3 K.  Solid lines are the best-fit results using the two-component function described in Eq. (1).}
\label{fig:Figure4}
\end{center}
\end{figure}

Finally, we discuss the comparison between hole-doped and electron-doped cuprates briefly in term of the Cu-spin correlation. 
From the $\mu$SR results of Zn-substituted La$_{2-x}$Sr$_x$Cu$_{1-y}$Zn$_y$O$_4$ in the overdoped regime, it has been found that the Zn-induced development of the Cu-spin correlation becomes weak with hole doping and finally disappears at $x \sim 0.30$ where the superconductivity disappears in LSCO. \cite{risdi-lsco} 
Moreover, inelastic neutron scattering experiments have uncovered the disappearance of the AF spin fluctuations concomitant with the disappearance of superconductivity at $x=0.30$. \cite{wakimoto} 
In the electron-doped cuprate, a short-range magnetic order due to a very small amount of excess oxygen is formed in the parent SC polycrystal of La$_{1.8}$Eu$_{0.2}$CuO$_{4+\delta}$ and the SC underdoped single crystal of Pr$_{1.3-x}$La$_{0.7}$Ce$_x$CuO$_4$ with $x=0.10$. \cite{adachi-review,adachi-jpsj2}
The Cu-spin correlation is moderately developed in the overdoped PLCCO with $x=0.14$ \cite{risdi-plcco} and the development of the Cu-spin correlation almost disappears in the heavily overdoped PLCCO with $x=0.20$ where the superconductivity disappears. 
Therefore, there is a similarity for both hole-doped and electron-doped cuprates in terms of the Cu-spin correlation, that is, the development of the Cu-spin correlation is observed in the SC region of the phase diagram. 
It is suggested that the Cu-spin correlation is in intimate relation with the appearance of high-$T_{\rm c}$ superconductivity in both hole- and electron-doped cuprates.

\section{Summary}
ZF-and LF-$\mu$SR spectra have revealed that the development of the Cu-spin correlation weakens with increasing $x$ and is negligibly small in heavily overdoped PLCCO with $x=0.20$ where the superconductivity disappears. 
These results suggest that the Cu-spin correlation exists in the electron-doped T'-cuprates where the superconductivity appears. 
It is suggested that, in both hole-doped and electron-doped cuprates, the mechanism of the superconductivity is related to the Cu-spin correlation.

\section*{Acknowledgments}
We would like to thank M. Ishikuro of the Institute for Materials Research, Tohoku University, Japan, for his help in the ICP analysis. 
This work was supported by JSPS KAKENHI Grant Numbers 23540399, 16K05458, 17H02915 and by MEXT KAKENHI Grant Number 23108004.


\end{document}